# An Autoethnographic Case Study of Generative Artificial Intelligence's Utility for Accessibility

Kate Glazko
Paul G. Allen School of Computer Science & Engineering, University of Washington, Seattle
Seattle, Washington, USA

Momona Yamagami
Paul G. Allen School of Computer Science & Engineering, University of Washington, Seattle
Seattle, Washington, USA

Aashaka Desai
Paul G. Allen School of Computer Science & Engineering, University of Washington, Seattle
Seattle, Washington, USA

Kelly Avery Mack
Paul G. Allen School of Computer Science & Engineering, University of Washington, Seattle
Seattle, Washington, USA

Venkatesh Potluri
Paul G. Allen School of Computer Science & Engineering, University of Washington, Seattle
Seattle, Washington, USA

Xuhai Xu
The Information School, University of Washington, Seattle
Seattle, Washington, USA

Jennifer Mankoff
jmankoff@acm.org
Paul G. Allen School of Computer Science & Engineering, University of Washington, Seattle
Seattle, Washington, USA

## Abstract

With the recent rapid rise in Generative Artificial Intelligence (GAI) tools, it is imperative that we understand their impact on people with disabilities, both positive and negative. However, although we know that AI in general poses both risks and opportunities for people with disabilities, little is known specifically about GAI in particular. To address this, we conducted a three-month autoethnography of our use of GAI to meet personal and professional needs as a team of researchers with and without disabilities. Our findings demonstrate a wide variety of potential accessibility-related uses for GAI while also highlighting concerns around verifiability, training data, ableism, and false promises.

## 1 Introduction

In the past few years, Generative Artificial Intelligence (GAI) [16] has seen exponential growth in the variety and power of tools available, especially after the release of ChatGPT in 2022 [7, 48]. However, concerns have been raised about the potential of the GAI to negatively impact inclusion, representation[2, 11, 42], and equity [26] for those in marginalized communities, including people with disabilities [13, 17, 18, 20]. This work has shown that GAI has ableist biases [17] [20] and may associate some forms of disabilities with toxic content [20] or harmful stereotypes [13]. Studies of other types of AI and machine learning have raised concerns about AI ethics "failing to attend to disability"[21] and shown mixed value (*e.g.,* [21, 27]). The release of and general availability of artificial intelligence (AI) and machine learning (ML) technologies in everyday devices and settings also creates new opportunities to understand how disabled people themselves approach and use these technologies, and the barriers they face when doing so. A few studies have explored how people with disabilities choose to use AI technologies in the field. For example, two studies of everyday use of home hubs have shown a mix of accessibility benefits and frustrations [37, 41]. Other work in participatory design has shown a disconnect between the current research de facto of building tools to increase capabilities, and the preferences of people with disabilities, with participants expressing desire for tools to help with social communication and interaction instead of productivity [30].

As GAI is being taken up by everyday users in a wide variety of domains from communication [1, 19, 34] to computer programming [35], it is imperative that we understand how GAI can be helpful and useful specifically to disabled people. While it has been proposed that GAI could be used to help individuals with visual impairment [9, 45] or autism [15], first-person accounts of GAI use in the field have the potential to provide a perspective grounded in the current capabilities, and flaws, of such technology. As GAI is rapidly adopted and embedded in existing tools and workflows, it is urgent that we explore the potential benefits and challenges posed by GAI. To address this gap in understanding, we came together as a team of seven individuals with and without disabilities to conduct a three-month autoethnography of our use of GAI to meet personal and professional needs. We used GAI to meet various access needs in domains including summarization, communication, image generation, GUI and visualization design, and making documents and visualizations accessible. We used currently available tools for these tasks, such as ChatGPT and Midjourney, and found that their utility varied significantly and almost always required human verification. We discuss concerns about verifiability and success metrics, the relevance of training data for some tasks, false promises, and the subtle nature of ableism in some of the GAI-generated results. It is crucial that GAI research directly tackles these challenges from a disability perspective or we are likely to see people with disabilities once again left out of the use and products of the newest generation of technology and communication tools.

## 2 Team and Methods

Our team is comprised of seven accessibility researchers with mixed abilities (including individuals who identify as blind, hard-of-hearing, neurodiverse, chronic illness, along with those who do not identify as disabled), different research experience levels (junior PhD students, senior PhD students, postdocs, senior faculty), minority backgrounds (diverse genders, LGBTQ+ statuses, immigration status, first-generation college student). Our team members had varied experiences with accessibility, including making artifacts accessible, teaching accessibility, prior training in accessibility, having taken one or more courses, and little to no training. In addition to GAI, our team members regularly use digital accessibility tools such as text-to-speech, screen magnifiers, automated captions and meeting transcripts/recordings and organizational or reminder applications; physical access tools such as a cane or noise-cancelling headphones; and modified work practices. Some of us had between 6 months and 1.5 years of prior experience using AI-powered accessibility technology to support our access needs prior to the start of data collection for this article in March 2023. The tools we used during this time included SeeingAI, Envision, DALL-E 2, ChatGPT, automatic image descriptions and automated captioning.

**Data Collection** Data collection started in March 2023 and continued until the first week of June 2023. During that time, we actively looked for opportunities to use generative AI tools to either address an access need for a disability or to make our research and teaching materials more accessible. During that time, various of us used Github Copilot, MidJourney, DALL-E 2, GPT4, GPT3, ChatPDF, and Bing to address various needs including accessible programming, communication, summarizing PDFs, formatting references, text simplification, image generation, and making PDFs screenreader accessible. We focused on opportunities to use GAI to directly solve a specific and personally relevant need.

During the data collection period, we each independently summarized our experiences in a shared document. We described our motivation for using generative AI to address our access needs, and noted what went well, what didn't go well, whether or not the generative AI results addressed the access need, and whether the results raised any issues related to ableism or representation. We also met weekly to discuss our process and findings.

**Analysis** At the end of the data collection period, each authors' data was coded by a second author, who noted themes in categories including: GAI's value for access, GAI's ability to produce accessible artifacts, the accessibility need being solved, and concerns raised. We discussed questions that arose, and then regrouped to discuss the themes once all of the coding was done. For anonymity we present our results using illustrative vignettes, amalgams of the data we collected. Two authors collaborated to develop these vignettes. All quotes are drawn directly from our autoethnographic notes.

## 3 Results

We categorize our experiences into two different types of needs: (1) Using GAI to help meet our own access needs (2) Using GAI to help make things more accessible for others. Within each, we group data with a common purpose into vignettes, such as assisting with interpersonal communication (*e.g., Vignetted 2: Sam*) or improving accessibility of visualization color choices (*e.g., Vignetted 7: Ruby*).

### 3.1 Can Generative AI Help Meet Access Needs for People with Disabilities?

In this section, we describe four different scenarios in which we consolidated our experiences as we attempted to use GAI to meet an access need present in our work, research, or daily lives. We describe the problems and barriers we encounter, describe our motivations for utilizing GAI as a possible solution, and detail the specifics of the GAI tools being used and how we used them.

**3.1.1 Summarization and Information Extraction** Information extraction is a common task of information workers, and may range from deep reading and understanding to searching for key facts such as authorship or date of publication. Brain fog, often associated with chronic illness can make processing text content difficult, [23]. This can impact everything from understanding or shortening text to more mundane tasks such as reference formatting, and motivated *Vignetted vignette:mia: .*

> **Vignette 1: Mia (amalgam of 2 people's experiences)**
>
> Mia, who has intermittent brain fog due to a chronic illness, chose to use GAI for "*…related work tasks because I don't have a good way to skim papers, and abstracts vary in how useful they are…it would be useful if I could take a paper and get a summary about it / ask specific questions about it*". They tried asking ChatPDF.com to summarize and answer specific questions about a paper. They reflected: "*Every once in a while [it] nailed an answer …[but] often gave me completely incorrect answers.*" In one example, they describe it

> *...Mia (amalgam of 2 people's experiences) cont.*
>
> incorrectly saying something was not present that was; in another they report a subtler and more insidious mistake. In a summary of a paper they were very familiar with, it oversimplified a nuanced argument, so that it "*[sounded like the authors were] suggesting talking to caregivers as proxies **instead of** chronically ill people . . . in the design process, as 'they can provide valuable insights into the needs and experiences of those with chronic illnesses (ChatPDF)'*" This mistake turns a finding from an award winning paper into an ableist and problematic trope (i.e., that you should talk to proxies over disabled people).
>
> Later that day, Mia began entering content into a paper they were writing, but their brain fog was increasing, making it hard to extract detailed information from web pages as was necessary to create references for some of the not-conventionally-academic content they were interacting with. They used GPT-4 and Bing to generate correctly formatted references that they could paste into their document. Mia described the experience: "*I was able to generate syntactically correct BibTeX entries (for the most part). [AI] hallucinations and errors were an issue, and also it would sometimes generate the wrong entry type (@article vrs @misc, for example)*". However, they did not mind correcting the Bibtex because "*having a start and checking if it was correct was much easier*". Mia stated that they did not use GAI to help with ideation or written content creation, because correcting it to match their voice took longer than writing themself (noting they are an adept and fast writer already).

In this vignette, GAI yielded mixed results. Results were not accurate and in some cases, unusable. In addition to hallucinations [10], GAI replaced nuanced arguments with simplified, ableist tropes. GAI was most useful, despite its errors, in the relatively mundane and low-stakes task of reference generation because it replaced a cognitively harder task (generating new, correctly formatted references from a web page) with a less difficult task (checking a reference for errors in specific fields).

**3.1.2 Interpersonal Communication Support** Interpersonal communication is probably the most universal use case for digital technology–text in particular can further promote misunderstandings[24], for example due to the lack of social cues [32]. This can potentially amplify worries for neurodiverse people about how their communication will be understood [12, 44].

> **Vignette 2: Sam (an amalgam of 1 person's experiences)**
>
> Sam, who is autistic, experiences social anxiety, particularly around casual, written communication. This, combined with past misunderstandings of their Slack communications at work, has led Sam to feel their tone is wordy and anxious; and spend "*too much time figuring out if I should send the message or not*". Inspired by a lecture describing ChatGPT as "*being good at sounding confident*", they decided to ask GPT4 to rewrite the messages to reflect "*confidence and [a] concise tone*". The results were "*written in a way I wish I could have

> *...Sam (an amalgam of 1 person's experiences) cont.*
>
> written them originally,*" leading Sam to feel "*immense relief*". Unsure whether the improvement was real, Sam tested for grammar improvements. Grammarly built-in statistics including a (1) numerical Readability score and (2) engagement and delivery reports showed that the revised messages were clearer and more engaging than the originals. However, when Sam shared both sets of messages with trusted peers and asked them to honestly assess which messages they liked better, they preferred the original in many cases and described the updated messages as "*robotic*" and "*not like a person wrote them*". Despite this mixed feedback, Sam still preferred the GPT-modified messages because being able to "*socially rubber duck with an AI made [them] feel more confident than writing alone*".
>
> Later, based on a recommendation from a peer who had success using ChatGPT to communicate with his non-English speaking parents, Sam decides to see if ChatGPT could do a better job of capturing emotional context in translation than Google Translate. Sam understands the language of their overseas relatives fairly well, but is not good at writing or speaking it. Due to Sam's autism, they already struggle with getting emotional content such as greetings, jokes, and cultural sayings right in their native language: Communicating with their foreign relatives through Google Translate often results in unfriendly, overly-formal translations with literal sayings that often don't make sense and correcting them requires a lot of modification- a cognitive-intensive task that Sam does not always have the bandwidth for. Sam asked ChatGPT to translate a friendly holiday greeting. The result "*isn't overly formal*" and uses friendly language and conventional sayings, an improvement over Google Translate.

This vignette highlights the limitations of current metrics and approaches for assessing the usefulness of AI generated text for improving written communication. Metrics, as well as the opinions of both communication partners are all important; and in any given circumstance the prioritization of these might need to vary.

**3.1.3 Visual Imagery Creation** Visual imagery creation, the translation of words or a concept into an image, is a common task useful in domains as broad as reading fiction for pleasure, designing amiguri (a Japanese crochet craft [31]) and generating illustrations and diagrams. Visual imagery creation requires a wide variety of cognitive ability, fine motor control, and artistic training or aptitude. This can make the creation of visual imagery inaccessible without proper support to people with a wide variety of impairments.

> **Vignette 3: Ally (amalgam of 2 people's experiences)**
>
> Ally, who has aphantasia (an inability to visualize noted for its impacts on the process of imagination [3]), was reading a passage in one of her favorite fiction novels and found herself "*once again skipping over a section describing scenery*" due to her aphantasia. She loves watching movie adaptations of her favorite books because they "*do the visualizing for

> **...Ally (amalgam of 2 people's experiences) cont.**
>
> *her*", but now has the opportunity to see such images on demand. She tries DALL-E 2 and Midjourney. In both cases, she enters a passage from the book describing the house in which the main character grew up. Dall-E 2 captured some elements of the ambience of the scene, but didn't perfectly match the image description; Midjourney is much better. Ally describes her excitement–she can finally "*see for herself what the scenes and characters in her favorite book could look like!*"
>
> Ally's inability to visualize extends beyond scenery in fiction books–it affects her ability to visually imagine new designs for crafts as well. Currently, she has to sketch them out on a piece of paper or use photoshop in order to "*get a visual*" of her concept, but her motor impairment has impacted her ability to draw or use a stylus–making both take longer. Ally uses Midjourney to generate photorealistic sketches of a novel craft concept and is "*thrilled*" by the results.
>
> 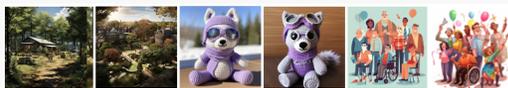
>
> **Figure 1: Midjourney visualizations, from left to right: Two scenery visualizations from a fiction novel, two prototype sketches of a crocheted lavender husky wearing ski goggles, two images of a group of people with a variety of disabilities looking happy.**
>
> Although an experienced UX developer, Ally's aphantasia and motor impairments also negatively impact her product design and development process, when she needs to generate images for others to look at. This is made worse when a migraine adds pain to the mix, exacerbated by the detail work and concentration graphic design requires. Ally needs to generate an illustration that includes a group of disabled people for a proposal. She prompts Midjourney: "*group of people with a variety of disabilities looking happy, illustration format*". The resulting images are not representative of many disabilities and "*inaccurate [in their] use of assistive technologies*" even after specific requests in new prompts. Also, despite repeated attempts to vary the prompt, Midjourney can "*only represent disabled happiness as attending a party*". In the end, she cannot produce any images of high enough quality to include in the proposal.

In this vignette, GAI again yielded mixed results, primarily when asked to envision anything to do with disability. In addition to its ableism, it is possible that the final example failed in part because the prompter had a preconceived image of what they wanted. Even an image that matched the prompt might easily be judged as in the wrong style, having the wrong background, or the wrong types of disabilities represented. Thus, some of the (unsuccessful) prompt iteration Ally went through might be necessary even without ableist results. At a certain point, if prompt iteration is lengthy enough, it might outweigh the cost of another method (e.g., Googling photos of groups of disabled people).

### 3.1.4 Graphical User Interface Design and Development
Interface design and development and data science are two of the most common computer programming jobs [6]. However, unlike other aspects of computer programming, Graphical User Interface / Web (GUI) design and development, and visualization design, are not easily accessible to screen reader users. Screen reader users can include people who experience nausea, migraine or exacerbation of other symptoms when using a screen, or people who are blind/low vision.

> **Vignette 4: Jay (amalgam of 1 person's experiences)**
>
> Jay is a legally blind, experienced software developer who uses a screenreader. He hopes GAI can help with GUI and visualization design, an area that current tools do not address well [36]. He describes one frustration, "*that visual semantics (size, shape and esthetics) of templates, assets, and other resources are often inaccessible*". Because he wants to build his own version of a stock-trading application, Jay asks ChatGPT with GPT4 to describe the general iconography used in IOS, and specific visual elements of the iOS default Stocks app (layout and colors) and to "*describe the layout of the stocks app*" and element ordering and color. Jay stated: "*Though it may not be the most updated, the iconography information [of the ios UI] and the description of the [layout of the stocks app] was super helpful to have.. I have some sense of the correctness of some of the icons based on my conversations with other sighted people.*". This provided enough information that Jay asked ChatGPT with GPT4 to provide starter Swift code based on the Stocks app layout. He says "*SwiftUI code generated seems correct based on my understanding of the language*". He however had no way to verify the correctness of the layout.
>
> Jay also wanted to include some advanced features into his app. He was exploring the feasibility of using latent dirichlet allocation (LDA), a topic modeling approach that can group words in a text corpus into topics. For the analysis, he "*tried to generate a data visualization to show results from a hyperparameter sweep to identify the right set of parameters to train [the] LDA model*". Creating an appropriate data visualization is time-consuming and laborious for Jay, because information about "*what specific data visualization function calls do, related documentation, and examples*" is frequently inaccessible.
>
> He used Github CoPilot with VSCode for this task: "*saved LDA results into a CSV file. Loaded it into a pandas dataframe, and added prompts as python comments to get GitHub Copilot to suggest matplotlib code to plot the graph. I tried to (1) accept the inline suggestions that the extension provided, and (2) got the extension to synthesize 10 suggestions and tried to make sense of the code to pick what I thought was the best. Went online and looked up documentation for the specific function calls that the extension was suggesting in either case*". While Jay described finding starter code as helpful, when he shared the visualization with a colleague, they "*very honestly said that the graph didn't make any sense at all*".

In this vignette, GAI provided some value to Jay, particularly when providing on-demand descriptions of an existing GUI. However, creating an interface or visualization was more problematic. The iterative loop of generating a visual artifact, looking at it, and updating code until it is closer to the goal state is very hard to engage when a second person is required for verification of the visual artifact.

### 3.2 Can Generative AI Help Make Content Accessible for Others?

In this section, we describe three different scenarios which consolidate our experiences of using GAI to make content that is accessible for others. As a result, we do not mention the disability of the creator in these cases.

**3.2.1 Document Accessibility** Document accessibility is a continual area of friction for people with disabilities, often made worse by tools that hide or even make it impossible to add accessibility meta data.

> **Vignette 5: Kaia (amalgam of 2 people's experiences)**
>
> Kaia is writing a research paper in LaTeX due to the preferences of her collaborators. She hopes GAI could help with other tasks such as making tables accessible. She explains that " *the easiest thing to do is to add lines between each cell and that is the easiest way to get it to tag correctly, so I was trying to see whether GPT can get there. I was also curious if it could make me a table that has lines that separate each cell but look nice visually as well.*" GPT-4 correctly described the steps necessary to make an accessible table: "*It knew that you can't make tables screen reader accessible on LaTeX and that you'd need to use Adobe Acrobat Pro to finish the task. The suggested steps were accurate. The original table it made, while not very pretty, would have been easy to convert to a screen reader accessible table*". Kaia asked GPT-4 to beautify the table, which failed. "*When I asked it to make the table look nice, it took away the lines that separate each cell which would be an issue when tagging the PDF later on. I tried to ask it to make the table look nice and then add thin lines later to separate the cells and …it failed [because] it used a package that actually does exist, but the command does not [exist]*"

> **Vignette 6: Jun (amalgam of 1 person's experiences)**
>
> Jun is a student preparing to teach a class as an instructor for the first time. He uses an online platform to prepare teaching materials and create slides together with another instructor collaboratively. Jun knows "*some general guidelines on how to make accessible slides. But I am not sure whether my slides check all boxes, especially because I am using a new online tool. Some underlying features may cause the slides to be inaccessible.*" He first used GPT-4 for general tips. These tips informed some aspects that Jun was not aware of, such as adding descriptions to all links and buttons via `aria-label`. Next, he used the online browsing version of GPT-4 to evaluate his specific slides. GPT-4 was able to identify some meta-information about the slides, such as the slide deck title and the creator. However, its responses on accessibility were very generic, for example, it suggested providing alternative text for images and closed captions for videos (even though the slides did not include a video). However, when he exported the slides in HTML format and inserted them at the end of the prompt, GPT-4 was able to highlight

> **…Jun (amalgam of 1 person's experiences) cont.**
>
> concrete accessibility issues, such as missing `alt` attribute for `navbar-toggler-icon`, inaccessible color contrast, and lacking a `type` attribute for a form input. However, Jun noticed that even in this case ChatGPT missed several problems, even after multiple prompts. Some of the problems were even just mentioned in the general guidelines that ChatGPT provided in the first step (*e.g.,* many buttons still didn't have the `aria-label` attribute).

In both of these examples, GAI was able to *parrot back accessibility rules*, but had a harder time executing them, especially while meeting other expectations such as aesthetic goals. Specific critiques were not comprehensive or consistent with general accessibility guidelines it had provided.

**3.2.2 Making Web Visualizations Accessible** Visualization accessibility is increasingly important in today's context of pandemic tracking; climate change and other domains where visualization is used to convey key points and support decision making (*e.g.,* [39, 40]). However, existing tools do not easily support the creation of accessible visualizations [22, 25, 28], especially to people who have color vision deficiency (CVD).

> **Vignette 7: Ruby (amalgam of two people's experiences)**
>
> 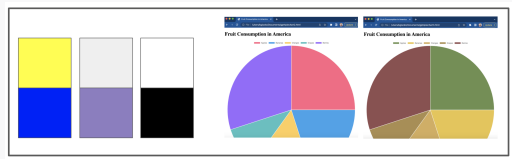
>
> Figure 2: GAI-generated colors and visualizations, from left to right: Color palettes recommended by GPT-4 including yellow-blue, lavendar-gray, and black white; and two different GPT-4 generated data visualizations, the first prompted as accessible, and the second prompted as CVD-friendly, with the latter consisting of amber, green, and yellow tones.
>
> Ruby is a Computer Science student who is tasked with designing data visualizations of readily-available datasets through D3 for their data visualization class. Ruby is not a design expert, stating: "*I hate picking colors and I don't know how to do it*". However, Ruby knows that they want colors that are nice *and* accessible. They asked GPT-4 to provide a color combination for a figure as inspiration. GPT-4 provided a commonly-used "*classic blue and yellow combo*". Iterating and specifying specific combinations- such as a purple and gray combination also yielded desirable results. However, GPT-4 also offered questionable aesthetic advice, highly recommending black and white as CVD-friendly colors, stating "*This is the most contrasted pair possible, and it is suitable for all forms of color blindness. Moreover, they are naturally printer-friendly*".

> *...Ruby (amalgam of two people's experiences) cont.*
>
> Given the partial success of this initial attempt, Ruby used GPT-4 again to help with their next assignment- this time, asking GPT-4 to produce the code for an accessible visualization of fruit consumption in the United States. GPT-4 correctly generated a visualization using ChartJS, but despite the request to make it accessible, it did not use accessible colors. Next, Ruby prompted GPT-4 to make it CVD friendly. GPT-4 produced a chart that was primarily shades of yellow, amber, and green- "*I had a hunch they were not color-blind friendly*". They compared the before-and-after on a CVD online simulator and was disappointed that the output "*looked worse than the before*".

Overall, GAI showed only a shallow grasp of how to improve color accessibility. Likewise, they did not present CVD-inclusive colors by default when asked to generate a visualization. In contrast to another visualization automation tool, Tableau [4], which selects a CVD-friendly palette of orange and blue as their default, GPT-4 did not appear to take CVD-accessible design into consideration when generating visualizations unprompted. Given the availability of color palette analysis tools that can provide CVD-friendly colors [43] and GPT-4's self-described knowledge of accessible colors, GPT-4 could be missing an already-built opportunity to "build in" accessibility into their generated data visualizations. We discuss our takeaways in the next section.

## 4 Discussion

Narratives around GAI in popular media are varied and quickly evolving, ranging from those who suggest it will be the "greatest technology ever developed" [33], to those envisioning it as a solution to most modern day dilemmas, to those who caution against its use without safety and ethics in mind [14, 46]. Our experiences may explain why GAI is viewed with such mixed emotions: Many, if not most, of the needs we were addressing *could not be met with existing, non-GAI tools*. By imagining a future in which GAI *could meet these needs* we also highlight the degree to which existing practices are inaccessible.

Yet our experiences also showed that use of GAI is not a flawless, autonomous experience, but rather is sometimes useful and sometimes not. Even our most positive experiences, such as automatically generating citations or on-the-fly visualizations, required interventions to either "fix" the outputs post-hoc or iterate and adjust the given prompts. On-demand accessibility is a future we should strive to support. However, we must address serious concerns with viability, verifiability, and ableism to get there.

**Verifiability:** We found GAI to be useful *as a tool, requiring significant human involvement and iteration,* rather than an out-of-the-box end-to-end solution for our access needs. As such, it was most useful when its answers were easy to verify (parts of *Vignetted 1: Mia* and *Vignetted 2: Sam*) or low stakes and exploratory. For instance, open-ended creative tasks that did not require a strict solution (parts of *Vignetted 3: Ally*) were more successful than trying to generate an image with specific, preconceived attributes such as disabled representation and assistive technology devices. GAI was least useful to us when *verification required the same or a similar accommodation GAI was being used to provide*, such as for paper summarization in *Vignetted 1: Mia* and visualization design in *Vignetted 4: Jay*. It is essential that we develop *complementary accessible representations* in these situations to better support verification. For example, occasionally creating a tactile version of a visualization, diagram, or user interface could help to support verification during the design process; this could complement going from text to visual design and back to text again and help support validation. Another alternative is to develop accessible metrics, summary statistics (such as readability in *Vignetted 2: Sam*). Metrics are much faster than direct investigation of the validity of results, but it is essential we clearly connect them to disabled preferences. For example, in *Vignetted 2: Sam*, Grammarly's metrics seemed to match Sam's preferences, while "preference of communication partner" did not. The development, and testing, of such metrics is still an under-explored space for GAI's use in on-demand access tasks.

**Relevant Training Data** While GAI did not generate complete solutions in any of our documented experiences, some results required many more prompt iterations than others. We suspect this may be due to the lack of relevant training data for some tasks. For example, converting an online post to a properly formatted reference (*Vignetted 1: Mia*) is a task that people with and without disabilities are equally likely to do, which makes it likely that there is at least *some* training data readily available in data sets that are not collected with accessibility in mind. In contrast, converting a visualization to a verbal description (or vice versa) (*Vignetted 4: Jay*) is not necessarily a common task, it is likely to show up only in accessible visualizations and presentations. Other tasks may depend on missing meta data – for example, a visualization designed with accessible colors for someone with CVD (*Vignetted 7: Ruby*) may not be labeled as such. Without better training data, these tasks are likely to remain inaccessible even as GAI quality advances.

**Practical Guidance:** We had to figure out how to use GAI for access as we went. Although GAI is "end user friendly", successful use depends on an understanding of what tools to use for what task, and how to design prompts that get the best result, which is not an intuitive process for end users [47]. Although there is an increasing amount of guidance available for GAI use in general, this same level of information is not currently available for accessibility needs. Better information about tools and prompts that promote access would be valuable.

**Built-in Ableism:** Ableism was subtle, but very much present in our experiences. Examples include the elision of key content in *Vignetted 1: Mia* and the difficulty of producing a representative and accurate image of disabled people in *Vignetted 3: Ally*. These biases certainly reflect biases in training data. However, such problems are made even more insidious by the way in which they are likely to confirm, rather than correct human biases, raising questions about how to ensure such mistakes are caught even during human verification.

**False Promises** When asked to make artifacts accessible, while stating confidently that it understood accessibility guidelines, GAI would repeatedly provide a false solution (*e.g., Vignetted 6: Jun, Vignetted 7: Ruby*). Although these errors could often be found "at a glance" by the user or a collaborator, these errors can sometimes be neglected without hands-on verification by a person with accessibility expertise. A less-experienced user could incorporate recommendations under the false belief that they solved an accessibility issue. Further research is needed to assess how GAI can self-verify or at least facilitate verification rather than simply proclaiming success.

## 5 Conclusion and Future Work

Our work demonstrates the potential for GAI today to be used by people with disabilities to provide on-demand support for their accessibility needs, in low-stakes, easily verifiable contexts. Unsurprisingly, some of the limitations we encountered reflect GAI's current ability to *parrot* [5] information without truly integrating it. This can explain why it struggled, for example, to translate accessibility guidelines into documents consistently (*Vignetted 6: Jun*). We hope that current research will improve these types of errors over time. However, our work also uncovered problems that require directed ongoing attention to fix, such as ableist errors; accessible and complementary approaches to verification; and better training data. The subtleties of some of these errors raise questions about how and whether it is possible to address, or even detect them automatically.

Future work should move beyond single-case explorations of GAI's capabilities to more systematic evaluations. In addition, there are additional on-demand needs that we did not test. For example, we believe that GAI has promise for on demand text simplification but did not explore this in depth or in the range of application domains this could be important. Ongoing research efforts are needed to design GAI algorithms, datasets, and experiences that are representative, are not ableist, and meet the needs of disabled individuals. Relatedly GAI is increasingly being used to produce artifacts that may or may not be accessible. For example, GAI is increasingly being used to write code [8, 35]. It is imperative to evaluate and improve the accessibility of such artifacts. If not, as GAI replaces other means of production, GAI could potentially move app and web accessibility *backwards* from the current, already problematic state of the world (*e.g.,* [29, 38]).

To summarize, we have presented an auto-ethnographic case study of real-world use of a variety of GAI technologies in the field by people with and without disabilities for a wide variety of tasks. Our results highlight the importance of careful experiments documenting the risks, and opportunities for GAI as well as new technical solutions for improving GAI's ability to support access.

## Acknowledgments

This work was funded by Meta, Center for Research and Education on Accessible Technology and Experiences (CREATE), Google, a NIDILRR ARRT Training Grant 90ARCP0005-01-00, the NSF GRFP under Grant No. DGE-2140004, and NSF EDA 2009977. Kate Glazko was supported by an an NSF CSGrad4US Graduate Fellowship and the UW Paul G. Allen School of Computer Science and Engineering Richard Ladner Endowed Fund for Graduate Student Support. Venkatesh Potluri was supported by the Apple Scholars in AI/ML PhD fellowship.